\documentclass[12pt]{article}
\usepackage{amsmath}
\usepackage{amssymb}
\usepackage{amsfonts}

\renewcommand{\theequation}{\arabic{section}.\arabic{equation}}
\newcommand{\field}[1]{\mathbb{#1}}
\newcommand{\R}{\field{R}}
\DeclareMathOperator{\Res}{Res}

\title{Extended Nikiforov-Uvarov method, roots of polynomial solutions, and functional Bethe ansatz method}
\author{C. Quesne\thanks{Electronic mail: cquesne@ulb.ac.be}\\ 
{\small\sl Physique Nucl\'eaire Th\'eorique et Physique Math\'ematique,  Universit\'e Libre de Bruxelles,} \\ 
{\small\sl Campus de la Plaine CP229, Boulevard~du Triomphe, B-1050 Brussels, Belgium}}
\date{ }
\begin{document}
\baselineskip=22pt plus 1pt minus 1pt
\maketitle
\begin{abstract}
For applications to quasi-exactly solvable Schr\"odinger equations in quantum mechanics, we establish the general conditions that have to be satisfied by the coefficients of a second-order differential equation with at most $k+1$ singular points in order that this equation has particular solutions which are $n$th-degree polynomials. In a first approach, we extend the Nikiforov-Uvarov method, which was devised to deal with hypergeometric-type equations (i.e., for $k=2$), and show that the conditions involve $k-2$ integration constants. In a second approach, we consider the functional Bethe ansatz method in its most general form. Comparing the two approaches, we prove that under the assumption that the roots of the polynomial solutions are real and distinct, the $k-2$ integration constants of the extended Nikiforov-Uvarov method can be expressed as linear combinations of monomial symmetric polynomials in those roots, corresponding to partitions into no more than two parts. 
\end{abstract}

\noindent
Keywords: Schr\"odinger equation, quasi-exactly solvable potentials, symmetric polynomials

\noindent
PACS Nos.: 03.65.Fd, 03.65.Ge
%
%
\newpage
\section{INTRODUCTION}

In quantum mechanics, solving the Schr\"odinger equation is a fundamental problem for understanding physical systems. Exact solutions may be very useful for developing a constructive perturbation theory or for suggesting trial functions in variational calculus for more complicated cases. However, very few potentials can actually be exactly solved (see, e.g., one of their lists in \cite{cooper}) \cite{footnote}. These potentials are connected with second-order differential equations of hypergeometric type and their wavefunctions can be constructed by using the theory of corresponding orthogonal polynomials \cite{szego}. Among the many methods used to deal with such cases, one may quote that of Nikiforov and Uvarov \cite{nikiforov}, which enables to solve systematically any generalized hypergeometric-type equation in order to obtain eigenvalues and eigenfunctions.\par
%
%
Apart from exactly solvable Schr\"odinger equations, the so-called quasi-exactly solvable (QES) ones, for which only a finite number of eigenstates can be found explicitly by algebraic means, while the remaining ones remain unknown, are also very interesting. The simplest QES problems, discovered in the 1980s, are characterized by a hidden sl(2,$\R$) algebraic structure \cite{turbiner87, turbiner88, ushveridze, gonzalez, turbiner16} and are connected with polynomial solutions of the Heun equation \cite{ronveaux}. Generalizations of this equation are related through their polynomial solutions to more complicated QES problems. In such a context, the functional Bethe ansatz (FBA) method \cite{gaudin, ho, zhang} has proven very effective \cite{agboola12, agboola13, agboola14, cq16}.\par
%
%
In some recent works, Karayer, Demirhan, and B\"uy\"ukk\i l\i \c c proposed an extension of the Nikiforov-Uvarov method to solve second-order differential equations, which have at most four singular points. These include the Heun and confluent Heun equations \cite{karayer15a}, as well as the biconfluent and triconfluent Heun equations \cite{karayer15b}. In addition, they demonstrated the usefulness of their method by explicitly solving some QES problems.\par
%
%
The purpose of the present paper is twofold: first to formulate the extended Nikiforov-Uvarov (ENU) method in its full generality to deal with second-order differential equations that have at most $k+1$ singular points, and second to establish a connection of the extended method with the FBA one.\par
%
%
In Section~II, we review the ENU method and show that the reduced equation that can be derived has particular solutions that are polynomials of degree $n$, depending on $k-2$ integration constants. On the assumption that such polynomials have real and distinct roots $z_1, z_2, \ldots, z_n$, we prove in Section~III that the corresponding integration constants satisfy a system of linear equations whose coefficients can be written in terms of elementary symmetric polynomials in $z_1, z_2, \ldots, z_n$ and we conjecture an explicit solution of this system in terms of monomial symmetric polynomials in $z_1, z_2, \ldots, z_n$. After deriving the FBA method in its most general form in Section~IV, in Section~V we provide a proof of the conjectured expression of the $k-2$ integration constants of the ENU method by comparing its results with those of the FBA one. Finally, Section~VI contains the conclusion.\par
%
%
\section{EXTENDED NIKIFOROV-UVAROV METHOD}

The starting point of the Nikiforov-Uvarov method \cite{nikiforov} is the second-order differential equation
\begin{equation}
  \psi''(z) + \frac{\tilde{\tau}(z)}{\sigma(z)} \psi'(z) + \frac{\tilde{\sigma}(z)}{\sigma^2(z)} \psi(z) = 0,
  \label{eq:NU-eq}
\end{equation}
where $\tilde{\tau}(z)$ is a polynomial of at most first degree, $\sigma(z)$ and $\tilde{\sigma}(z)$ are polynomials of at most second degree, and $\psi(z)$ is a function of hypergeometric type. The criteria related to degrees of polynomial coefficients constitute the so-called boundary conditions of the method. To deal with solutions of Heun-type equations, Karayer, Demirhan, and B\"uy\"ukk\i l\i \c c changed such boundary conditions in such a way that $\tilde{\tau}(z)$, $\sigma(z)$, and $\tilde{\sigma}(z)$ became polynomials of at most second, third, and fourth degree, respectively.\par
%
%
In the present approach, we will assume that in Eq.~(\ref{eq:NU-eq}), $\tilde{\tau}(z)$, $\sigma(z)$, and $\tilde{\sigma}(z)$ are polynomials of at most $(k-1)$th, $k$th, and $(2k-2)$th degree, respectively. By setting
\begin{equation}
  \psi(z) = \phi(z) y(z),
\end{equation}
where $\phi(z)$ is some suitable function, which will be determined later on, Eq.~(\ref{eq:NU-eq}) is converted to
\begin{equation}
  y''(z) + \left(2 \frac{\phi'(z)}{\phi(z)} + \frac{\tilde{\tau}(z)}{\sigma(z)}\right) y'(z) + \left(\frac{\phi''(z)}{\phi(z)}
  + \frac{\phi'(z)}{\phi(z)} \frac{\tilde{\tau}(z)}{\sigma(z)} + \frac{\tilde{\sigma}(z)}{\sigma^2(z)}\right) y(z) = 0.
  \label{eq:diff-y}
\end{equation}
\par
%
%
Such an equation can be simplified by rewriting the coefficients of $y'(z)$ and $y(z)$ in terms of some newly defined polynomials. For the coefficient of $y'(z)$, we take
\begin{equation}
  2 \frac{\phi'(z)}{\phi(z)} + \frac{\tilde{\tau}(z)}{\sigma(z)} = \frac{\tau(z)}{\sigma(z)},  \label{eq:tau}
\end{equation}
where $\tau(z)$ is a polynomial of at most $(k-1)$th degree, and, in addition, we set
\begin{equation}
  \tau(z) = \tilde{\tau}(z) + 2\pi(z)  \label{eq:pi}
\end{equation}
in terms of a polynomial $\pi(z)$ of at most $(k-1)$th degree. On combining (\ref{eq:tau}) with (\ref{eq:pi}), we get
\begin{equation}
  \frac{\phi'(z)}{\phi(z)} = \frac{\pi(z)}{\sigma(z)}.  \label{eq:phi}
\end{equation}
Furthermore, for the coefficient of $y(z)$ in (\ref{eq:diff-y}), we set
\begin{equation}
  \frac{\phi''(z)}{\phi(z)} + \frac{\phi'(z)}{\phi(z)} \frac{\tilde{\tau}(z)}{\sigma(z)} + \frac{\tilde{\sigma}(z)}{\sigma^2(z)}
  = \frac{\bar{\sigma}(z)}{\sigma^2(z)}, 
\end{equation}
where $\bar{\sigma}(z)$ is a polynomial of at most $(2k-2)$th degree, given by
\begin{equation}
  \bar{\sigma}(z) = \tilde{\sigma}(z) + \pi^2(z) + \pi(z) [\tilde{\tau}(z) - \sigma'(z)] + \pi'(z) \sigma(z).
  \label{eq:sigma-bar}
\end{equation}
Equation (\ref{eq:diff-y}) therefore becomes
\begin{equation}
  y''(z) + \frac{\tau(z)}{\sigma(z)} y'(z) + \frac{\bar{\sigma}(z)}{\sigma^2(z)} y(z) = 0.
\end{equation}
\par
%
%
If the polynomial $\bar{\sigma}(z)$ is divisible by $\sigma(z)$, i.e.,
\begin{equation}
  \frac{\bar{\sigma}(z)}{\sigma(z)} = h(z),  \label{eq:h}
\end{equation}
where $h(z)$ is a polynomial of degree at most $k-2$, then we get a reduced equation
\begin{equation}
  \sigma(z) y''(z) + \tau(z) y'(z) + h(z) y(z) = 0.  \label{eq:red-eq}
\end{equation}
On using definition (\ref{eq:h}) in (\ref{eq:sigma-bar}) and setting
\begin{equation}
  h(z) - \pi'(z) = g(z),  \label{eq:g}
\end{equation}
which is a polynomial of degree at most $k-2$, we get a quadratic equation for the polynomial $\pi(z)$, namely
\begin{equation}
  \pi^2(z) + [\tilde{\tau}(z) - \sigma'(z)] \pi(z) + \tilde{\sigma}(z) - g(z) \sigma(z) = 0.
\end{equation}
Its roots are given by
\begin{equation}
  \pi(z) = \tfrac{1}{2} [\sigma'(z) - \tilde{\tau}(z)] \pm \Bigl\{\tfrac{1}{4} [\sigma'(z) - \tilde{\tau}(z)]^2
  - \tilde{\sigma}(z) + g(z) \sigma(z)\Bigr\}^{1/2}.  \label{eq:eq-pi}
\end{equation}
\par
%
%
To determine all possible solutions for the polynomial $\pi(z)$ from Eq.~(\ref{eq:eq-pi}), the polynomial $g(z)$ under the square root sign must be known explicitly. Since $\pi(z)$ is a polynomial of degree at most $k-1$, the expression under the square root sign must be the square of a polynomial of degree at most $k-1$. There are in general several possibilities for choosing $g(z)$ in such a way that the latter condition is satisfied. For every of them, two solutions for the polynomials $\pi(z)$ can be obtained from (\ref{eq:eq-pi}). Then $\tau(z)$, $h(z)$, and $\phi(z)$ can be found from Eqs.~(\ref{eq:pi}), (\ref{eq:g}), and (\ref{eq:phi}), respectively.\par
%
%
To be really useful, the solutions of the reduced equation (\ref{eq:red-eq}) have to be generalized. On deriving this equation $k-2$ times, we obtain
\begin{align}
  & \sigma y^{(k)} + \left[\binom{k-2}{1} \sigma' + \tau\right] y^{(k-1)} + \left[\binom{k-2}{2} \sigma'' + \binom{k-2}{1}
       \tau' + h \right] y^{(k-2)} + \cdots \nonumber \\
  & + \left[\sigma^{(k-2)} + \binom{k-2}{1} \tau^{(k-3)} + \binom{k-2}{2} h^{(k-4)}\right] y'' + \left[\tau^{(k-2)} +
       \binom{k-2}{1} h^{(k-3)}\right] y' \nonumber \\
  & + h^{(k-2)} y = 0,  \label{eq:diff-red-eq} 
\end{align}
which is a $k$th-order differential equation with polynomial coefficients of degree not exceeding the corresponding order of differentiation. Since all its derivatives have the same form, it can be differentiated $n$ times by using the new representation $y^{(n)}(z) = v_n(z)$. In such a notation, Eq.~(\ref{eq:diff-red-eq}) can be written as
\begin{align}
  & \sigma v_0^{(k)} + \left[\binom{k-2}{1} \sigma' + \tau\right] v_0^{(k-1)} + \left[\binom{k-2}{2} \sigma'' 
       + \binom{k-2}{1} \tau' + h \right] v_0^{(k-2)} + \cdots \nonumber \\
  & + \left[\sigma^{(k-2)} + \binom{k-2}{1} \tau^{(k-3)} + \binom{k-2}{2} h^{(k-4)}\right] v_0'' + \left[\tau^{(k-2)} +
       \binom{k-2}{1} h^{(k-3)}\right] v_0' \nonumber \\
  & + h^{(k-2)} v_0 = 0. 
\end{align}
Its $n$th derivative can be easily shown to be given by
\begin{align}
  & \sum_{l=0}^k \binom{n+k-2}{k-l} \sigma^{(k-l)} v_n^{(l)} + \sum_{l=0}^{k-1} \binom{n+k-2}{k-l-1} \tau^{(k-l-1)}
       v_n^{(l)} \nonumber \\
  & + \sum_{l=0}^{k-2} \binom{n+k-2}{k-l-2} h^{(k-l-2)} v_n^{(l)} = 0.
\end{align}
When the coefficient of $v_n$ in this equation is equal to zero, i.e.,
\begin{equation}
  \binom{n+k-2}{k} \sigma^{(k)} + \binom{n+k-2}{k-1} \tau^{(k-1)} + \binom{n+k-2}{k-2} h^{(k-2)} = 0,
  \label{eq:condition}
\end{equation}
there exists a particular solution $y(z) = y_n(z)$ that is a polynomial of degree $n$. On integrating Eq.~(\ref{eq:condition}) $k-2$ times, we find that this occurs whenever $h(z) = h_n(z)$ is given by
\begin{equation}
  h_n(z) = - \frac{n(n-1)}{k(k-1)} \sigma''(z) - \frac{n}{k-1} \tau'(z) + \sum_{l=0}^{k-3} C_{k-l-2,n} \frac{z^l}{l!},
  \label{eq:h_n}
\end{equation}
where $C_{1,n}, C_{2,n}, \ldots, C_{k-2,n}$ are $k-2$ integration constants.\par
%
%
It is worth observing here that in the Nikiforov-Uvarov hypergeometric case \cite{nikiforov}, we have $k=2$ so that the polynomial $h(z)$ reduces to a constant $\lambda$. Then $\lambda = \lambda_n$, where, in accordance with Eq.~(\ref{eq:h_n}),
\begin{equation}
  \lambda_n = - \tfrac{1}{2} n(n-1) \sigma''(z) - n \tau'(z),
\end{equation}
with no integration constant. In the Heun-type equation case of Refs.~\cite{karayer15a, karayer15b}, we have $k=3$ and the linear polynomial $h(z) = h_n(z)$ is given by
\begin{equation}
  h_n(z) = - \tfrac{1}{6} n(n-1) \sigma''(z) - \tfrac{1}{2} n \tau'(z) + C_n
\end{equation}
in terms of a single integration constant $C_n$ (see Eq.~(24) of \cite{karayer15a}).\par
%
%
At this stage, as in \cite{karayer15a, karayer15b}, we might assume some specific forms of $\tilde{\tau}(z)$, $\sigma(z)$, and $\tilde{\sigma}(z)$ for some $k$ and determine all types of polynomial solutions of the reduced equation (\ref{eq:red-eq}) that can be obtained by selecting all allowed $g(z)$ in (\ref{eq:eq-pi}) and setting $h(z) = h_n(z)$. Instead of doing this, in Section~III we will proceed to interpret the $k-2$ integration constants $C_{i,n}$, $i=1, 2, \ldots, k-2$, of Eq.~(\ref{eq:h_n}) in terms of the roots of the polynomial solutions $y_n(z)$.\par
%
%
\section{INTEGRATION CONSTANTS AND ROOTS OF POLYNOMIAL SOLUTIONS}

\setcounter{equation}{0}

In this Section and the following ones, we slightly change the notations used in Section~II and rewrite the reduced equation (\ref{eq:red-eq}) as
\begin{equation}
  X(z) y''(z) + Y(z) y'(z) + Z(z) y(z) = 0,  \label{eq:red-eq-bis}
\end{equation}
where
\begin{equation}
  X(z) = \sum_{l=0}^k a_l z^l, \qquad Y(z) = \sum_{l=0}^{k-1} b_l z^l, \qquad Z(z) = \sum_{l=0}^{k-2} c_l z^l,
  \label{eq:X-Y-Z}
\end{equation}
and $a_l$, $b_l$, $c_l$ are some (real) constants.\par
%
%
In Section~II, we have shown that $n$th-degree polynomial solutions $y_n(z)$ of Eq.~(\ref{eq:red-eq-bis}) can be obtained provided $Z(z)$ is given by
\begin{equation}
  Z_n(z) = - \frac{n(n-1)}{k(k-1)} X''(z) - \frac{n}{k-1} Y'(z) + \sum_{l=0}^{k-3} C_{k-l-2,n} \frac{z^l}{l!},
  \label{eq:Z_n}
\end{equation}
where $C_{1,n}, C_{2,n}, \ldots, C_{k-2,n}$ are some integration constants. On inserting Eq.~(\ref{eq:X-Y-Z}) in (\ref{eq:Z_n}) and equating the coefficients of equal powers of $z$ on both sides, we obtain the set of relations
\begin{align}
  c_{k-2} &= - n(n-1) a_k - n b_{k-1}, \label{eq:rel-1} \\
  c_l &= - \frac{n(n-1)}{k(k-1)} (l+2)(l+1) a_{l+2} - \frac{n}{k-1} (l+1) b_{l+1} + \frac{C_{k-l-2,n}}{l!}, \nonumber \\
  & \qquad l=0, 1, \ldots, k-3.  \label{eq:rel-2}
\end{align}
\par
%
%
The $n$th-degree polynomial solutions $y_n(z)$ of the reduced equation (\ref{eq:red-eq-bis}) can be written as
\begin{equation}
  y_n(z) = \prod_{i=1}^n (z-z_i),  \label{eq:roots}
\end{equation}
where we assume that the roots $z_1, z_2, \ldots, z_n$ are real and distinct. We now plan to show that the integration constants $C_{1,n}, C_{2,n}, \ldots, C_{k-2,n}$ satisfy a system of linear equations whose coefficients can be expressed in terms of elementary symmetric polynomials in $z_1, z_2, \ldots, z_n$ \cite{littlewood},
\begin{equation}
\begin{split}
  e_l &\equiv e_l(z_1, z_2, \ldots, z_n) = \sum_{1\le i_1< i_2< \cdots < i_l\le n} z_{i_1} z_{i_2} \ldots z_{i_l}, \qquad 
      l=1, 2, \ldots, n, \\
  e_0 &\equiv 1.
\end{split}  \label{eq:e_l}
\end{equation}
\par
%
%
We can indeed rewrite $y_n(z)$ in (\ref{eq:roots}) as
\begin{equation}
  y_n(z) = \sum_{m=0}^n (-1)^{n-m} e_{n-m} z^m,
\end{equation}
so that
\begin{align}
  y'_n(z) &= \sum_{m=0}^{n-1} (-1)^{n-m-1} (m+1) e_{n-m-1} z^m, \\
  y''_n(z) &= \sum_{m=0}^{n-2} (-1)^{n-m-2} (m+2)(m+1) e_{n-m-2} z^m.
\end{align}
On inserting these expressions in Eq.~(\ref{eq:red-eq-bis}) and taking Eq.~(\ref{eq:X-Y-Z}) into account, we get
\begin{align}
  & \sum_{l=0}^k a_l z^l \sum_{m=0}^{n-2} (-1)^{n-m-2} (m+2)(m+1) e_{n-m-2} z^m \nonumber \\
  & + \sum_{l=0}^{k-1} b_l z^l \sum_{m=0}^{n-1} (-1)^{n-m-1} (m+1) e_{n-m-1} z^m \nonumber \\
  & + \sum_{l=0}^{k-2} c_l z^l \sum_{m=0}^n (-1)^{n-m} e_{n-m} z^m = 0.  \label{eq:red-eq-ter}
\end{align}
\par
%
%
Here $l+m$ runs from 0 to $k+n-2$. Let us therefore set $l+m = k+n-r$, where $r=2, 3, \ldots, k+n$. Equation (\ref{eq:red-eq-ter}) can then be rewritten as
\begin{align}
  & \sum_{r=2}^k \left\{\sum_{p=0}^{r-2} (-1)^p \left[a_{k-r+2+p} (n-p)(n-p-1) + b_{k-r+1+p} (n-p) + c_{k-r+p}\right]
       e_p\right\} z^{k+n-r} \nonumber \\
  & \qquad + \text{(lower-degree terms with $k+1 \le r \le k+n$)} = 0.
\end{align}
On setting to zero the coefficients of $z^{k+n-r}$, $r=2, 3, \ldots, k$, we obtain the relations
\begin{align}
  & \sum_{p=0}^{r-2} (-1)^p \left[a_{k-r+2+p} (n-p)(n-p-1) + b_{k-r+1+p} (n-p) + c_{k-r+p}\right] e_p = 0, \nonumber \\
  & \qquad r=2, 3, \ldots, k.  
\end{align}
For $r=2$, we simply get $n(n-1) a_k + n b_{k-1} + c_{k-2} = 0$, which is automatically satisfied due to Eq.~(\ref{eq:rel-1}). We are therefore left with the $k-2$ relations
\begin{align}
  & \sum_{p=0}^{r-3} (-1)^p \left[a_{k-r+2+p} (n-p)(n-p-1) + b_{k-r+1+p} (n-p) + c_{k-r+p}\right] e_p \nonumber \\
  & + (-1)^{r-2} [a_k (n-r+2)(n-r+1) + b_{k-1} (n-r+2) + c_{k-2}] e_{r-2} = 0, \nonumber \\
  & \qquad r=3, 4, \ldots, k.
\end{align} 
\par
%
%
After substituting the right-hand sides of Eqs.~(\ref{eq:rel-1}) and (\ref{eq:rel-2}) for $c_{k-2}$ and $c_{k-r+p}$ in these relations, we obtain a system of $k-2$ linear equations for the $k-2$ integration constants $C_{1,n}, C_{2,n}, \ldots, C_{k-2,n}$,
\begin{align}
 & \sum_{p=0}^{r-3} (-1)^p \frac{C_{r-2-p,n}}{(k-r+p)!} e_p = - \sum_{p=0}^{r-3} (-1)^p \biggl\{\frac{1}{k(k-1)} 
     \Bigl[(r-p-2)(2k-r+p+1) n^2 \nonumber \\
 &\quad - [2pk^2 + 2k(r-2p-2) - (r-p-2)(r-p-1)] n + k(k-1)p(p+1)\Bigr] a_{k-r+2+p} \nonumber \\
 &\quad + \frac{1}{k-1} [(r-p-2) n - (k-1)p] b_{k-r+1+p}\biggr\} e_p \nonumber \\
 &\quad - (-1)^{r-1} (r-2) [(2n-r+1) a_k + b_{k-1}] e_{r-2}, \qquad r=3, 4, \ldots, k.  \label{eq:syst-C}
\end{align}
\par
%
%
The determinant of this system having zeros above the diagonal is easily determined to be given by $\left[\prod_{r=3}^k (k-r)!\right]^{-1} \ne 0$. It is therefore obvious that the constants $C_{1,n}, C_{2,n}, \ldots , C_{k-2,n}$ can be calculated successively from the equations corresponding to $r=3, 4, \ldots, k$.\par
%
%
It turns out that instead of elementary symmetric polynomials $e_l(z_1, z_2, \ldots, z_n)$, defined in Eq.~(\ref{eq:e_l}), it is more appropriate to express the solution in terms of monomial symmetric polynomials in $z_1, z_2, \ldots, z_n$, 
\begin{equation}
  m_{(\lambda_1, \lambda_2, \ldots, \lambda_n)}(z_1, z_2, \ldots, z_n) = \sum_{\pi \in S_{\lambda}} 
  z_{\pi(1)}^{\lambda_1} z_{\pi(2)}^{\lambda_2} \ldots z_{\pi(n)}^{\lambda_n},
\end{equation}
where $(\lambda_1, \lambda_2, \ldots, \lambda_n)$ denotes a partition and $S_{\lambda}$ is the set of permutations giving distinct terms in the sum \cite{littlewood}. The derivation of the first three constants $C_{1,n}$, $C_{2,n}$, and $C_{3,n}$ is outlined in the Appendix. In particular, it is shown there that $m_{(1^3,\dot{0})}$ (a dot over zero meaning that it is repeated as often as necessary), corresponding to a partition into more than two parts and which in principle might appear in $C_{3,n}$, actually does not occur because it has a vanishing coefficient. This is a general property that we have observed for the first six constants that we have computed explicitly and which all agree with the general formula
\begin{align}
  \frac{C_{q,n}}{(k-2-q)!} &= - \sum_{t=0}^{q-1} [2(n-1) a_{k-t} + b_{k-t-1}] m_{(q-t,\dot{0})} - \sum_{s=1}^{[q/2]}
       \sum_{t=0}^{q-2s} 2a_{k-t}\, m_{(q-t-s,s,\dot{0})} \nonumber \\
  &\quad - \frac{n(n-1)}{k(k-1)} q(2k-q-1) a_{k-q} - \frac{n}{k-1} q b_{k-q-1}, \nonumber \\
  &\qquad q=1, 2, \ldots, k-2,  \label{eq:C}
\end{align}
where $[q/2]$ denotes the largest integer contained in $q/2$.\par
%
%
At this stage, Eq.~(\ref{eq:C}) is a conjecture, which might be proved from (\ref{eq:syst-C}) by determining $C_{q,n}$ for any $q \in \{1, 2, \ldots, k-2\}$. This would, however, be a rather complicated derivation. In Section~V, we will proceed to show that a much easier proof of (\ref{eq:C}) can be found by comparing the results of the ENU method with those of the FBA one.\par
%
%
\section{FUNCTIONAL BETHE ANSATZ METHOD}

\setcounter{equation}{0}

In its most general form, the FBA method also starts from the reduced equation (\ref{eq:red-eq-bis}), with $X(z)$, $Y(z)$, $Z(z)$ given in (\ref{eq:X-Y-Z}), and considers polynomial solutions of type (\ref{eq:roots}) with real and distinct roots $z_1, z_2, \ldots, z_n$ \cite{gaudin, ho, zhang}. Equation (\ref{eq:red-eq-bis}) is then rewritten as
\begin{equation}
  - c_0 = \left(\sum_{l=0}^k a_l z^l\right) \sum_{i=1}^n \frac{1}{z-z_i} \sum_{\substack{
  j=1 \\
  j\ne i}}^n \frac{2}{z_i-z_j} + \left(\sum_{l=0}^{k-1} b_l z^l\right) \sum_{i=1}^n \frac{1}{z-z_i} + \sum_{l=1}^{k-2}
  c_l z^l.  \label{eq:constant}
\end{equation}
The left-hand side of this equation is a constant, while the right-hand one is a meromorphic function with simple poles at $z=z_i$ and a singularity at $z=\infty$. Since the residues at the simple poles are given by
\begin{equation}
  \Res(-c_0)_{z=z_i} = \left(\sum_{l=0}^k a_l z_i^l\right) \sum_{\substack{
  j=1 \\
  j\ne i}}^n \frac{2}{z_i-z_j} + \sum_{l=0}^{k-1} b_l z_i^l, 
\end{equation}
Eq.~(\ref{eq:constant}) yields
\begin{align}
  - c_0 &= \sum_{l=1}^k a_l \sum_{i=1}^n \sum_{m=0}^{l-1} z_i^m z^{l-m-1}\sum_{\substack{
      j=1 \\
      j\ne i}}^n \frac{2}{z_i-z_j} + \sum_{l=1}^{k-1} b_l \sum_{i=1}^n \sum_{m=0}^{l-1} z_i^m z^{l-m-1} \nonumber \\
  &\quad + \sum_{l=1}^{k-2} c_l z^l + \sum_{i=1}^n \frac{\Res(-c_0)_{z=z_i}}{z-z_i}.
\end{align}
On defining 
\begin{equation}
  S_m = \sum_{i=1}^n \sum_{\substack{
      j=1 \\
      j\ne i}}^n \frac{z_i^m}{z_i-z_j}, \qquad T_m = \sum_{i=1}^n z_i^m,
\end{equation}
and observing that $S_0=0$, this relation becomes
\begin{equation}
  - c_0 = 2 \sum_{l=2}^k a_l \sum_{m=1}^{l-1} S_m z^{l-m-1} + \sum_{l=1}^{k-1} b_l \sum_{m=0}^{l-1} T_m z^{l-m-1}
  + \sum_{l=1}^{k-2} c_l z^l + \sum_{i=1}^n \frac{\Res(-c_0)_{z=z_i}}{z-z_i}, 
\end{equation}
or, with $q \equiv l-m-1$ in the first two terms and $q \equiv l$ in the third one,
\begin{equation}
  - c_0 = 2 \sum_{q=0}^{k-2} z^q \sum_{m=1}^{k-1-q} a_{q+m+1} S_m + \sum_{q=0}^{k-2} z^q \sum_{m=0}^{k-2-q}
       b_{q+m+1} T_m + \sum_{q=1}^{k-2} z^q c_q + \sum_{i=1}^n \frac{\Res(-c_0)_{z=z_i}}{z-z_i}. 
\end{equation}
\par
%
%
The right-hand side of this equation will be a constant if and only if the coefficients of $z^q$, $q=1, 2, \ldots, k-2$, and all the residues at the simple poles vanish. This yields $c_q$, $q=1, 2, \ldots, k-2$, in terms of the coefficients of $X(z)$, $Y(z)$, and the roots of $y_n(z)$,
\begin{equation}
  c_q = - 2 \sum_{m=1}^{k-1-q} a_{q+m+1} S_m - \sum_{m=0}^{k-2-q} b_{q+m+1} T_m, \qquad q=1, 2, \ldots, k-2,
  \label{eq:c_q}
\end{equation}
as well as the $n$ algebraic equations determining the roots, i.e., the Bethe ansatz equations,
\begin{equation}
  \sum_{\substack{
  j=1 \\
  j\ne i}}^n \frac{2}{z_i-z_j} + \frac{\sum_{l=0}^{k-1} b_l z_i^l}{\sum_{l=0}^k a_l z_i^l} = 0, \qquad i=1, 2, \ldots, n.
\end{equation}
The remaining constant leads to the value of $c_0$,
\begin{equation}
  c_0 = - 2 \sum_{m=1}^{k-1} a_{m+1} S_m - \sum_{m=0}^{k-2} b_{m+1} T_m.  \label{eq:c_0}
\end{equation}
\par
%
%
It remains to find the explicit expressions of $S_m$ and $T_m$. For the smallest allowed $m$ values, it is obvious that
\begin{equation}
  S_1 = \tfrac{1}{2} n(n-1), \qquad T_0 = n.
\end{equation}
For higher $m$ values, $S_m$ can be written as a linear combinations of monomial symmetric polynomials in $z_1, z_2, \ldots, z_n$. From
\begin{equation}
  S_m = \frac{1}{2} \sum_{i=1}^n \sum_{\substack{
  j=1 \\
  j\ne i}}^n \frac{z_i^m - z_j^m}{z_i-z_j} = \frac{1}{2} \sum_{i=1}^n \sum_{\substack{
  j=1 \\
  j\ne i}}^n \sum_{p=0}^{m-1} z_i^{m-1-p} z_j^p,
\end{equation}
we get for odd $m \ge 3$,
\begin{align}
  S_m &= \frac{1}{2} \sum_{i=1}^n \sum_{\substack{
      j=1 \\
      j\ne i}}^n \Biggl[z_i^{m-1} + z_j^{m-1} + \sum_{p=1}^{(m-3)/2} \bigl(z_i^{m-1-p} z_j^p + z_i^p z_j^{m-1-p}\bigr)
      + z_i^{(m-1)/2} z_j^{(m-1)/2}\Biggr] \nonumber \\
  &= (n-1) \sum_{i=1}^n z_i^{m-1} + \sum_{\substack{
      i, j=1 \\
      i \ne j}}^n \sum_{p=1}^{(m-3)/2} z_i^{m-1-p} z_j^p + \sum_{\substack{
      i, j=1 \\
      i < j}}^n z_i^{(m-1)/2} z_j^{(m-1)/2} \nonumber \\
  &= (n-1) m_{(m-1,\dot{0})} + \sum_{p=1}^{(m-1)/2} m_{(m-1-p,p,\dot{0})}, 
\end{align}
and for even $m \ge 2$,
\begin{align}
  S_m &= \frac{1}{2} \sum_{i=1}^n \sum_{\substack{
      j=1 \\
      j\ne i}}^n \Biggl[z_i^{m-1} + z_j^{m-1} + \sum_{p=1}^{(m-2)/2} \bigl(z_i^{m-1-p} z_j^p + z_i^p z_j^{m-1-p}\bigr)
      \Biggr] \nonumber \\
  &= (n-1) \sum_{i=1}^n z_i^{m-1} + \sum_{\substack{
      i, j=1 \\
      i \ne j}}^n \sum_{p=1}^{(m-2)/2} z_i^{m-1-p} z_j^p  \nonumber \\
  &= (n-1) m_{(m-1,\dot{0})} + \sum_{p=1}^{(m-2)/2} m_{(m-1-p,p,\dot{0})}. 
\end{align}
Hence, 
\begin{equation}
  S_m = (n-1) m_{(m-1,\dot{0})} + \sum_{p=1}^{[(m-1)/2]} m_{(m-1-p,p,\dot{0})}, \qquad m \ge 2.
\end{equation}
\par
%
%
{}Furthermore, it is obvious that
\begin{equation}
  T_m = m_{(m,\dot{0})}, \qquad m \ge 1.
\end{equation}
\par
%
%
On replacing $S_m$ and $T_m$ by their explicit values in (\ref{eq:c_q}) and (\ref{eq:c_0}), we get
\begin{align}
  c_{k-2} &= - n(n-1) a_k - n b_{k-1},  \label{eq:rel-1-bis} \\
  c_l &= - n(n-1) a_{l+2} - n b_{l+1} - 2 \sum_{m=2}^{k-1-l} a_{l+m+1} \biggl[(n-1) m_{(m-1,\dot{0})} \nonumber \\
  &\quad + \sum_{p=1}^{[(m-1)/2]} m_{(m-1-p,p,\dot{0})}\biggr] - \sum_{m=1}^{k-2-l} b_{l+m+1} m_{(m,\dot{0})},
       \nonumber \\
  &\qquad l=0, 1, \ldots, k-3.  \label{eq:rel-2-bis}
\end{align}
\par
%
%
\section{COMPARISON BETWEEN THE ENU AND FBA METHODS}

\setcounter{equation}{0}

Direct comparison between Eqs.~(\ref{eq:rel-1}), (\ref{eq:rel-2}) and Eqs.~(\ref{eq:rel-1-bis}), (\ref{eq:rel-2-bis}) shows that $c_{k-2}$ is given by the same expression in both methods, while for $l=0, 1, \ldots, k-3$, $c_l$ is written in terms of $a_{l+2}$ and $b_{l+1}$, as well as the integration constant $C_{k-l-2,n}$ in the ENU method or a linear combination of monomial symmetric polynomials in $z_1, z_2, \ldots, z_n$ in the FBA one.\par
%
%
Equating the two expressions for $c_l$, $l=0, 1, \ldots, k-3$, yields
\begin{align}
  \frac{C_{k-2-l,n}}{l!} &= - 2 \sum_{m=2}^{k-1-l} a_{l+m+1} \biggl[(n-1) m_{(m-1,\dot{0}} + \sum_{p=1}^{[(m-1)/2]}
       m_{(m-1-p,p,\dot{0})}\biggr] \nonumber \\
  &\quad - \sum_{m=1}^{k-2-l} b_{l+m+1} m_{(m,\dot{0})} - \frac{n(n-1)}{k(k-1)} (k-l-2)(k+l+1) a_{l+2} \nonumber \\
  &\quad - \frac{n}{k-1} (k-l-2) b_{l+1}, \qquad l=0, 1, \ldots, k-3.  \label{eq:C-bis}
\end{align}
On setting $q=k-2-l$ in Eq.~(\ref{eq:C-bis}), the latter becomes
\begin{align}
  \frac{C_{q,n}}{(k-2-q)!} &= - 2 \sum_{m=2}^{q+1} a_{k+m-1-q} \biggl[(n-1) m_{(m-1,\dot{0}} + \sum_{p=1}^{[(m-1)/2]}
       m_{(m-1-p,p,\dot{0})}\biggr] \nonumber \\
  &\quad - \sum_{m=1}^{q} b_{k+m-1-q} m_{(m,\dot{0})} - \frac{n(n-1)}{k(k-1)} q(2k-q-1) a_{k-q} \nonumber \\
  &\quad - \frac{n}{k-1} q b_{k-q-1}, \qquad l=0, 1, \ldots, k-3,  \label{eq:C-ter}
\end{align} 
where we see that the last two terms on the right-hand side coincide with the corresponding ones in Eq.~(\ref{eq:C}). The other terms can also be easily converted into those of Eq.~(\ref{eq:C}) by changing the summation indices. With $t=q+1-m$ and $t=q-m$, we can indeed rewrite
\begin{equation}
  \sum_{m=2}^{q+1} a_{k+m-1-q} m_{(m-1,\dot{0})} = \sum_{t=0}^{q-1} a_{k-t} m_{(q-t,\dot{0})}
\end{equation}
and
\begin{equation}
  \sum_{m=1}^q b_{k+m-1-q} m_{(m,\dot{0})} = \sum_{t=0}^{q-1} b_{k-t-1} m_{(q-t,\dot{0})},
\end{equation}
respectively. Furthermore, $t=q+1-m$ and $s=p$ lead to
\begin{align}
  & \sum_{m=2}^{q+1} a_{k+m-1-q} \sum_{p=1}^{[(m-1)/2]} m_{(m-1-p,p,\dot{0})} \nonumber \\
  &= \sum_{t=0}^{q-1} a_{k-t} \sum_{s=1}^{[(q-t)/2]} m_{(q-t-s,s,\dot{0})} \nonumber \\
  &= \sum_{s=1}^{[q/2]} \sum_{t=0}^{q-2s} a_{k-t} m_{(q-t-s,s,\dot{0})}.
\end{align}
\par
%
%
Collecting all the terms shows that Eq.~(\ref{eq:C-ter}) coincides with Eq.~(\ref{eq:C}), which is therefore proved.\par
%
%
\section{CONCLUSION}

In the present paper, we have established the general conditions that have to be satisfied by the coefficients of a second-order  differential equation with at most $k+1$ singular points in order that the equation has particular solutions that are $n$th-degree polynomials $y_n(z)$. This has been done in two different ways.\par
%
%
In the first one, we have extended the Nikiforov-Uvarov method \cite{nikiforov}, which was devised to deal with hypergeometric-type equations, i.e., for the $k=2$ case, and we have shown that the extended method involves $k-2$ integration constants. The generalization that we have proposed includes as a special case that considered by Karayer, Demirhan, and B\"uy\"ukk\i l\i \c c for Heun-type equations corresponding to $k=3$ \cite{karayer15a, karayer15b}.\par
%
%
In the second approach, we have presented the FBA method \cite{gaudin} in its most general form. Our results also include previous applications of the method \cite{ho, zhang, agboola12, agboola13, agboola14, cq16} as special cases.\par
%
%
Comparing the outcomes of both descriptions, we have proved that under the assumption that the roots $z_1, z_2, \ldots, z_n$ of the polynomial solutions $y_n(z)$ are real and distinct, the $k-2$ integration constants of the ENU method can be expressed as linear combinations of monomial symmetric polynomials in $z_1, z_2, \ldots, z_n$, corresponding to partitions into no more than two parts.\par
%
%
\section*{\boldmath APPENDIX: THE INTEGRATION CONSTANTS $C_{1,n}$, $C_{2,n}$, AND $C_{3,n}$}

\renewcommand{\theequation}{A.\arabic{equation}}
\setcounter{equation}{0}

The purpose of this Appendix is to solve Eq.~(\ref{eq:syst-C}) for $r=3$, 4, 5 and to show that the resulting expressions of $C_{1,n}$, $C_{2,n}$, and $C_{3,n}$ agree with Eq.~(\ref{eq:C}).\par
%
%
{}For $r=3$, Eq.~(\ref{eq:syst-C}) directly leads to
\begin{equation}
  \frac{C_{1,n}}{(k-3)!} = - [2(n-1) a_k + b_{k-1}] e_1 - \frac{2n(n-1)}{k} a_{k-1} - \frac{n}{k-1} b_{k-2},
  \label{eq:A1}
\end{equation}
which corresponds to Eq.~(\ref{eq:C}) for $q=1$ because $e_1 = m_{(1,\dot{0})}$.\par
%
%
{}For $r=4$, Eq.~(\ref{eq:syst-C}) becomes
\begin{align}
  & \frac{C_{2,n}}{(k-4)!} - \frac{C_{1,n}}{(k-3)!} e_1 \nonumber \\
  &= 2 [(2n-3) a_k + b_{k-1}] e_2 + \Bigl[\frac{2}{k}(n-1)(n-k) a_{k-1} + \frac{1}{k-1}(n-k+1) b_{k-2}\Bigr] e_1
       \nonumber \\
  &\quad - \frac{2n(n-1)}{k(k-1)} (2k-3) a_{k-2} - \frac{2n}{k-1} b_{k-3}.  \label{eq:A2}
\end{align}
On inserting (\ref{eq:A1}) in (\ref{eq:A2}) and using the identities $e_2 = m_{(1^2,\dot{0})}$, $e_1^2 = m_{(2,\dot{0})} + 2 m_{(1^2,\dot{0})}$, we get
\begin{align}
  \frac{C_{2,n}}{(k-4)!} &= - [2(n-1) a_k + b_{k-1}] m_{(2,\dot{0})} - [2(n-1) a_{k-1} + b_{k-2}] m_{(1,\dot{0})}
      \nonumber \\
  &\quad - 2a_k m_{(1^2,\dot{0})} - \frac{2n(n-1)}{k(k-1)} (2k-3) a_{k-2} - \frac{2n}{k-1} b_{k-3},  \label{eq:A3}
\end{align}
which agrees with Eq.~(\ref{eq:C}) for $q=2$.\par
%
%
On setting now $r=5$ in Eq.~(\ref{eq:syst-C}), we obtain
\begin{align}
  & \frac{C_{3,n}}{(k-5)!} - \frac{C_{2,n}}{(k-4)!} e_1 + \frac{C_{1,n}}{(k-3)!} e_2 \nonumber \\
  &= - 3[2(n-2) a_k + b_{k-1}] e_3 - \Bigl\{\frac{2}{k} [n^2 - (2k+1)n + 3k] a_{k-1} + \frac{1}{k-1} (n-2k+2) b_{k-2}
       \Bigr\} e_2 \nonumber \\
  &\quad + \Bigl\{\frac{2}{k(k-1)} (n-1)[(2k-3)n - k(k-1)] a_{k-2} + \frac{1}{k-1} (2n-k+1) b_{k-3}\Bigr\} e_1
       \nonumber \\
  &\quad - \frac{6n(n-1)}{k(k-1)} (k-2) a_{k-3} - \frac{3n}{k-1} b_{k-4}.  \label{eq:A4}
\end{align}
Here, let us employ Eqs.~(\ref{eq:A1}) and (\ref{eq:A3}), as well as the identities $e_3 = m_{(1^3,\dot{0})}$, $m_{(2,\dot{0})} m_{(1,\dot{0})} = m_{(3,\dot{0})} + m_{(2,1,\dot{0})}$, and $m_{(1^2,\dot{0})} m_{(1,\dot{0})} = m_{(2,1,\dot{0})} + 3 m_{(1^3,\dot{0})}$. For the coefficient of $m_{(1^3,\dot{0})}$ in $C_{3,n}/(k-5)!$, we obtain $- 3[2(n-2) a_k + b_{k-1}]$ from the right-hand side of (\ref{eq:A4}), $-6a_k$ from $C_{2,n} e_1/(k-4)!$, and $3[2(n-1) a_k + b_{k-1}]$ from $- C_{1,n} e_2/(k-3)!$, respectively. We conclude that $m_{(1^3,\dot{0})}$ does not occur in $C_{3,n}/(k-5)!$, which is given by
\begin{align}
  \frac{C_{3,n}}{(k-5)!} &= - [2(n-1) a_k + b_{k-1}] m_{(3,\dot{0})} - [2(n-1) a_{k-1} + b_{k-2}] m_{(2,\dot{0})}
       \nonumber \\
  &\quad - [2(n-1) a_{k-2} + b_{k-3}] m_{(1,\dot{0})} - 2a_k m_{(2,1,\dot{0})} - 2a_{k-1} m_{(1^2,\dot{0})}
       \nonumber \\
  &\quad - \frac{6n(n-1)}{k(k-1)} (k-2) a_{k-3} - \frac{3n}{k-1} b_{k-4},
\end{align}
in agreement with Eq.~(\ref{eq:C}) for $q=3$.\par
%
%


\newpage

\begin{thebibliography}{99}

\bibitem{cooper}
F.\ Cooper, A.\ Khare, and U.\ Sukhatme,
Phys.\ Rep.\ {\bf 251}, 267 (1995).

\bibitem{footnote}
Here we do not plan to discuss the recent development of the exceptional orthogonal polynomials and the associated polynomially solvable analytic potentials (see, e.g., \cite{gomez09, gomez10, cq08, cq09, odake09, odake11}).

\bibitem{szego}
G.\ Szeg\"o,
{\sl Orthogonal Polynomials} (American Mathematical Society, New York, 1939).

\bibitem{nikiforov}
A.\ V.\ Nikiforov and V.\ B.\ Uvarov,
{\sl Special Functions of Mathematical Physics} (Birkhauser, Boston, 1988).

\bibitem{turbiner87}
A.\ V.\ Turbiner and A.\ G.\ Ushveridze,
Phys.\ Lett.\ A {\bf 126}, 181 (1987).

\bibitem{turbiner88}
A.\ V.\ Turbiner,
Commun.\ Math.\ Phys.\ {\bf 118}, 467 (1988).

\bibitem{ushveridze}
A.\ G.\ Ushveridze,
{\sl Quasi-Exactly Solvable Models in Quantum Mechanics} (IOP, Bristol, 1994).

\bibitem{gonzalez}
A.\ Gonz\'alez-L\'opez, N.\ Kamran, and P.\ J.\ Olver,
Commun.\ Math.\ Phys.\ {\bf 153}, 117 (1993).

\bibitem{turbiner16}
A.\ V.\ Turbiner,
Phys.\ Rep.\ {\bf 642}, 1 (2016).

\bibitem{ronveaux}
A.\ Ronveaux,
{\sl Heun Differential Equations} (Oxford University Press, Oxford, 1995).

\bibitem{gaudin}
M.\ Gaudin,
{\sl La Fonction d'Onde de Bethe} (Masson, Paris, 1983).

\bibitem{ho}
C.-L.\ Ho,
Ann.\ Phys.\ {\bf 323}, 2241 (2008).

\bibitem{zhang}
Y.-Z.\ Zhang,
J.\ Phys.\ A: Math.\ Theor.\ {\bf 45}, 065206 (2012).

\bibitem{agboola12}
D.\ Agboola and Y.-Z.\ Zhang,
Mod.\ Phys.\ Lett.\ A {\bf 27}, 1250112 (2012).

\bibitem{agboola13}
D.\ Agboola and Y.-Z.\ Zhang,
Ann.\ Phys.\ {\bf 330}, 246 (2013).

\bibitem{agboola14}
D.\ Agboola, J.\ Links, I.\ Marquette, and Y.-Z.\ Zhang,
J.\ Phys.\ A: Math.\ Theor.\ {\bf 47}, 395305 (2014).

\bibitem{cq16}
C.\ Quesne,
Families of quasi-exactly solvable extensions of the quantum oscillator in curved spaces,
arXiv:1612.00682.

\bibitem{karayer15a}
H.\ Karayer, D.\ Demirhan, and F.\ B\"uy\"ukk\i l\i \c c,
J.\ Math.\ Phys.\ {\bf 56}, 063504 (2015).

\bibitem{karayer15b}
H.\ Karayer, D.\ Demirhan, and F.\ B\"uy\"ukk\i l\i \c c,
Rep.\ Math.\ Phys.\ {\bf 76}, 271 (2015).

\bibitem{littlewood}
D.\ E.\ Littlewood,
{\sl A University Algebra: An Introduction to Classic and Modern Algebra} (Dover, New York, 1971).

\bibitem{gomez09}
D.\ G\'omez-Ullate, N.\ Kamran, and R.\ Milson,
J.\ Math.\ Anal.\ Appl.\ {\bf 359}, 352 (2009).

\bibitem{gomez10}
D.\ G\'omez-Ullate, N.\ Kamran, and R.\ Milson,
J.\ Approx.\ Theory {\bf 162}, 987 (2010).

\bibitem{cq08}
C.\ Quesne,
J.\ Phys.\ A: Math.\ Theor.\ {\bf 41}, 392001 (2008).

\bibitem{cq09}
C.\ Quesne,
SIGMA {\bf 5}, 084 (2009).

\bibitem{odake09}
S.\ Odake and R.\ Sasaki,
Phys.\ Lett.\ B {\bf 679}, 414 (2009).

\bibitem{odake11}
S.\ Odake and R.\ Sasaki,
Phys.\ Lett.\ B {\bf 702}, 164 (2011).

\end{thebibliography}
\end{document}